# Interaction of terahertz electromagnetic field with metallic grating: Near-field zone

Lyaschuk Yu.M. Korotyeyev V.V.

Department of Theoretical Physics, V. Lashkaryov Institute of Semiconductor Physics,
National Academy of Sciences of Ukraine, 01650 Kyiv, Ukraine,
yulashchuk@gmail.com , koroteev@ukr.net .

**Abstract.** We have developed theory for the interaction of THz radiation with a sub-wavelength metallic grating. The structure of electric fields of the electromagnetic wave under the metallic grating has been studied in the near-field zone. Spatial distributions of the electric field components and the electric energy density have been obtained for the transmitted wave through the grating. An effect of strong local enhancement of the electric field has been detected. Spatial dependences of the polarization of the transmitted wave has been analyzed for the near-field zone.

**UDC** 537.874, 537.862

### 1. Introduction

In recent years, one of the priority directions in development of modern nanoelectronics and optoelectronics is elaboration of solid-state sources, detectors and modulators of electromagnetic (EM) waves in terahertz (THz) spectral range. This researches has been inspired by numerous potential applications of THz optics, including communication technologies, wireless local area networks, spectral analysis of complex molecules and materials, time-domain spectroscopy, THz imaging ( in particular, for medical applications), etc. [1, 2].

One of the most attractive directions in the development of THz optoelectronics is THz plasmonics that investigates controlling of the corresponding radiation by using excitation of plasma oscillations (plasmons) in semiconductor microdevices. During the past decade a lot of theoretical and experimental works have been devoted to problems of detecting and emitting the THz radiation by means of excitation of plasma waves in the 2D channel of the field-effect transistors [3, 4]. Here better operation conditions have been achieved for the transistors with multi-gated structures. A periodic system of metallic gates (i.e., a metallic grating) plays a role of specific antenna element that provides efficient coupling of relatively long-wavelength THz-range EM-radiation with short-wavelength plasmons. Resonant detection and amplification of THz radiations have recently been studied for the multi-gated field-effect transistors [5–8]. Moreover, similar multi-contact devices can be utilized as modulators and polarizers for the THz radiation [9, 10]. In experiments with transmission of the THz radiation one typically uses structures with a metallic grating of a sub-wavelength period ($a$) and submicron distance ($D$) between the grating and a 2D electron gas. Such a geometry is preferable for efficient excitation of plasmons with THz frequencies. The interaction of EM-waves with plasmons in these structures occurs in the near-field zone ($D << a$) near metallic gratings, where the properties of EM-field are very different from peculiar for the far-field zone.

The present article addresses a detailed analysis of structure of the EM-field transmitted through the single metallic grating in the near-field zone. In Section 2 we describe a theoretical approach employed for solving the problem of interaction of the EM-waves with the periodic metallic grating. The results of calculations for the spatial distribution of the EM-field located under the grating and its polarization characteristics are discussed in Section 3 and 4. Finally, the conclusions are drawn in Section 5.

**2. Theory of light interaction with metallic grating.**

A typical geometry of structure of grating is shown in Fig.1. The grating consists of metallic strips, with the width $b$ and the thickness $d$, which are arranged along the $x$ direction with the period $a$. The system is assumed to be a uniform and infinite along the $y$ direction. Let a plane monochromatic EM-wave be incident upon the grating along the $z$ axis with the electric vector polarized in the $x$ direction, (i.e., perpendicular to the grating strips). We consider the case of a subwavelength grating ($a < \lambda_0$), with $\lambda_0$ denoting the wavelength of the incident EM-wave. The grating strips will be treated as conducting layers infinitely thin along the $z$ direction. This assumption is justified whenever a skin layer is much thicker than $d$ (e.g., the estimations for the gold strips yield in the skin layer thickness of about 80 nm at 1 THz). In order to satisfy the assumption above mentioned, all of our calculations will be carried out for the $d = 20$ nm. Finally, the grating system is considered as being placed into uniform environment with the dielectric constant $\varepsilon$.

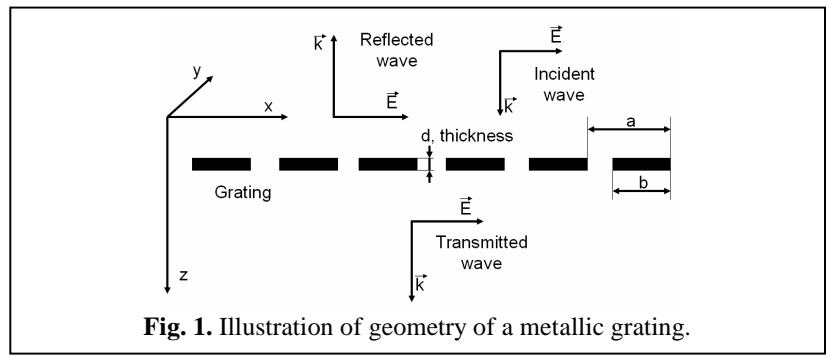

**Fig. 1.** Illustration of geometry of a metallic grating.

Being a result of interaction of the plane wave with the metallic grating, the total electric field obeys the Maxwell equation,

$$\text{rot rot } \vec{E}(\vec{r},t) + \frac{\varepsilon}{c^2}\frac{\partial^2 \vec{E}(\vec{r},t)}{\partial t^2} = -\frac{4\pi}{c^2}\frac{\partial \vec{j}(\vec{r},t)}{\partial t}. \qquad (1)$$

In Eq. (1), $\vec{j}(\vec{r},t)$ is the conduction current induced by the EM-field in the metallic strips. Due to symmetry of the problem, the electric field ($\vec{E}(r,t) = \{E_x(x,z), 0, E_z(x,z)\}\exp(-i\omega t)$) and the current ($\vec{j}(r,t) = \{j_x^{2D}(x)\delta(z), 0, 0\}\exp(-i\omega t)$) have two components and one component, respectively. Then the vector equation given by formula (1) may be rewritten as a system of two scalar equations:

$$\frac{\partial^2 E_z(x,z)}{\partial x \partial z} - \frac{\partial^2 E_x(x,z)}{\partial^2 z} - \omega^2 \frac{\varepsilon}{c^2} E_x = \frac{4\pi i \omega}{c^2} j_x^{2D}(x)\delta(z),$$

$$\frac{\partial^2 E_x(x,z)}{\partial x \partial z} - \frac{\partial^2 E_z(x,z)}{\partial^2 x} - \omega^2 \frac{\varepsilon}{c^2} E_z = 0. \qquad (2)$$

According to the Ohm law, the current $j_x(x)$ in the strips may be expressed in terms of the total electric field in the plane $z=0$:

$$j_x^{2D}(x) = \sigma^{2D}(x) E_x(x,0). \qquad (3)$$

with $\sigma^{2D}(x)$ being the local 2D conductivity of the metallic strips. Structural periodicity of the grating along the $x$ direction allows one searching for the solution of the system of Eqs. (2) in the form of Fourier series expansion:

$$E_{(x,z)}(x,z) = \sum_{m=-\infty}^{+\infty} \left\{ E_{x,m}^i(z) + E_{(x,z),m}^s(z) \right\} e^{i q_m x}, \qquad (4)$$

where $q_m$ is the wave number of the grating, ($q_m = 2\pi m/a$). Eq. (4) describes the total electric field as a sum of two contributions. The first one, $E_x^i(z)$, describes the external *incident* field. We choose this field to be a plane monochromatic wave ($E_{x,m}^i(z) = E^i \delta_{m,0} e^{i k_0 z}$, with $k_0 = \sqrt{\varepsilon}\omega/c$). The second contribution, $E_{(x,z)}^s(z)$, is the *scattered* field which is a result of re-emission induced by the alternating current produced in the metallic strips.

After inserting Eq. (4) into Eqs. (2) we obtain the equation for *m*th Fourier coefficient of the *x*-component of the scattered field, $E_{x,m}^s$:

$$\frac{\partial^2 E_{x,m}^s}{\partial z^2} - \lambda_m^2 E_{x,m}^s = \frac{4\pi i \lambda_m^2}{\varepsilon \omega} j_{x,m}^{2D}, \qquad (5)$$

along with the relationship between *m*th Fourier coefficients for the *x*- and *z*- components:

$$E^s_{z,m} = -i\frac{q_m}{\lambda_m^2}\frac{\partial E^s_{x,m}}{\partial z}. \tag{6}$$

Eq. (5) has the solution:

$$E^s_{m,x}(z) = A_m e^{\lambda_m z} \quad \text{at} \quad z < 0 \quad \text{and} \quad E^s_{m,x}(z) = B_m e^{-\lambda_m z} \quad \text{at} \quad z > 0, \tag{7}$$

where $\lambda_m = \sqrt{q_m^2 - k_0^2}$. When $m = 0$, we have $\lambda_m = -ik_0$ (here the sign minus was chosen in such that solutions would give a radiating modes) and the constants $A_0$ and $B_0 + E^i$ are amplitudes respectively of the reflected and the transmitted wave in the far-field zone.

In the case of sub-wavelength grating, $\lambda_m$ purely real at $m \neq 0$ and so the constants $A_m$ and $B_m$ represent the amplitudes of *evanescent* waves. The evanescent waves are localized near the grating and their intensity decreases exponentially with increasing distance from the grating. Due to translation symmetry of the grating along the x direction, the evanescent waves are equivalent for $q_m$ and $q_{-m}$. Therefore, the EM-field in the near field zone is nothing but a standing wave. The unknown constants $A_m$ and $B_m$ may be found from the following boundary conditions for plane $z = 0$:

$$E^s_{x,m}(z)|_{z=+0} = E^s_{x,m}(z)|_{z=-0}; \quad \frac{\partial E^s_{x,m}(z)}{\partial z}\bigg|_{z=+0} - \frac{\partial E^s_{x,m}(z)}{\partial z}\bigg|_{z=-0} = \frac{4\pi i \lambda_m^2}{\omega \varepsilon} j^{2D}_{x,m}. \tag{8}$$

The conditions given by Eqs. (8) imply that $A_m = B_m$ and $B_m = -2\pi i \lambda_m j^{2D}_{x,m}/\omega\varepsilon$. Using the Ohm law (see Eq. (3)), one can express the *m*th Fourier coefficient of the current as a convolution product of the Fourier components of the 2D conductivity and the total electric field calculated at $z = 0$:

$$j^{2D}_{x,m} = \sum_{m'=-\infty}^{\infty} \sigma^{2D}_{m-m'} E_{x,m'}\bigg|_{z=0}. \tag{9}$$

Taking into account that $B_m + E^i \delta_{m,0} = E_{x,m}|_{z=0}$ we get:

$$\sum_{m'=-\infty}^{\infty} \left\{ \delta_{m,m'} + \frac{2\pi i \lambda_m}{\varepsilon \omega} \sigma^{2D}_{m-m'} \right\} E_{x,m'}\bigg|_{z=0} = E^i \delta_{m,0} \tag{10}$$

Thus, we have reduced the electrodynamics problem of light interaction with the periodic metallic grating to an infinite system of algebraic equations. After solving numerically Eqs. (10), with large though finite number $M$ of the equations, one can obtain the spatial distribution of all the components of the electric field vector of the transmitted wave:

$$E_x(x,z) = \sum_{m=-M}^{M} E_{x,m}\bigg|_{z=0} e^{-\lambda_m z} e^{iq_m x}, \quad E_z(x,z) = \sum_{m=-M}^{M} \frac{iq_m}{\lambda_m} E_{x,m}\bigg|_{z=0} e^{-\lambda_m z} e^{iq_m x} \tag{11}$$

Good enough convergence of the solutions given by Eqs. (11) with increasing of number $M$

can be achieved while assuming that metallic strip is described by some smoothed conductivity profile. In the present work, we make use of the profile defined by

$$\sigma^{2D}(x) = \{\sigma_0^{2D}\sin^p(\pi x/b), \quad x\in[0,b] \quad \text{and} \quad 0, \quad x\in[b,a]\},$$

where $p$ means a fractional number. This profile is convenient because its Fourier coefficients have the analytical form:

$$\sigma_m^{2D} = \sigma_0^{2D} f 2^{1-p}(p+1)^{-1}\exp(-i\pi mf)B^{-1}(p/2+mf+1, p/2-mf+1). \tag{12}$$

with $f = b/a$ being the filling factor of the grating and $B$ the Euler beta-function.

Notice that another analytical approach to the grating-related problems is often used in the literature it. In frame of this method, the system of Eqs. (10) is reformulated in terms of an integral equation in the coordinate space. The general scheme of its solution is based on expansion of total field in the series of polynomials orthogonal with respect to the weight function $\sigma^{2D}(x)$. This procedure is again reduced to an infinite set of equations for the expansion coefficients. The approximate solution of the latter has somewhat better convergence when compared with our system (see Eqs. (10)). However, such a procedure can only be applied for specific profile shapes $\sigma^{2D}(x)$. For instance, the authors of Ref. [11] have analyzed plasma eigenmodes for the system consisting of grating and a 2D electron gas, using a semi-elliptic profile of the strip conductivity. Meanwhile our method is applicable for arbitrary profiles $\sigma^{2D}(x)$. Specific results obtained while solving directly Eqs. (10) are described in the next section.

### 3. Properties of the electric field under the grating

In Fig.2 we illustrate instantaneous spatial distribution of the $x$- and $z$- components of the total electric field $\text{Re}[E_{x,z}(x,z)e^{-i\omega t}]$ of the transmitted wave below the grating. The coordinates $x\in[0, 0.5a]$ and $x\in[0.5a, a]$ correspond to the location of the metallic strip and the window between the strips, respectively. As seen from Fig.2, the electric field has essentially non-uniform distribution in the near-field zone (see curves 1 and 2). This is a result of complicate superposition of the incident and scattered fields, the latter being induced redistribution of charge in the metallic strips.

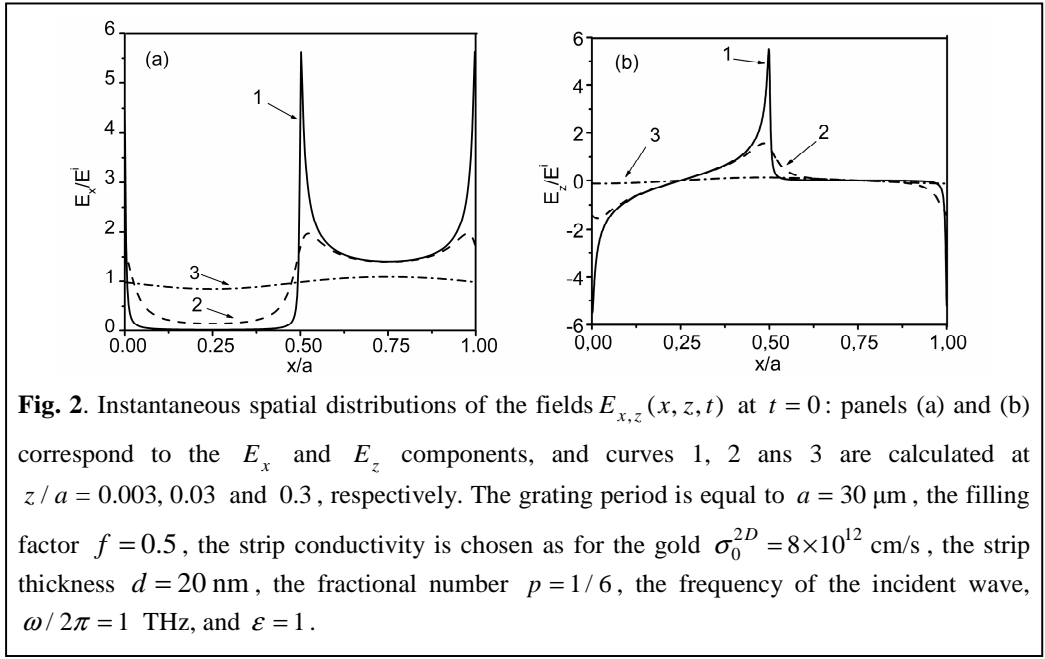

**Fig. 2**. Instantaneous spatial distributions of the fields $E_{x,z}(x,z,t)$ at $t=0$: panels (a) and (b) correspond to the $E_x$ and $E_z$ components, and curves 1, 2 ans 3 are calculated at $z/a = 0.003, 0.03$ and $0.3$, respectively. The grating period is equal to $a = 30$ μm, the filling factor $f = 0.5$, the strip conductivity is chosen as for the gold $\sigma_0^{2D} = 8\times10^{12}$ cm/s, the strip thickness $d = 20$ nm, the fractional number $p = 1/6$, the frequency of the incident wave, $\omega/2\pi = 1$ THz, and $\varepsilon = 1$.

Notice also that the scattered field is a superposition of the evanescent waves. One can observe a strong enhancement of electric field near the edges of the metallic strips. Both the x- and z- components exist in this narrow region and, moreover, each of them has the values much greater than the amplitude of the incident wave. It is worthwhile that the z- component is equal zero at the point $x = 0.25\,a$ (i.e., in middle of the strip). This fact indicates that spatial distribution of the scattered field looks like the field of a dipole with the corresponding electric charges induced on the opposite edges of the strip. The dipole has such an instantaneous polarity that its field almost totally screens the incident field under the stripe. It is clearly seen in Fig.2 that the total field has a very small x-component in the main region under the metallic strip, though the z-component remains non-vanishing, except for in the very middle of the strip.

However we observe opposite situation under the window (except for the edges regions). Here the incident field is not screened, but rather amplified by the scattered field, so that the former field mainly contributes to the total one. The total field has a vanishing z-component and a non-vanishing *x*-component, which remains practically uniform. The values of the *x*-component of the electric field are slightly larger than the amplitude of the incident wave. The spatial distribution of the total electric field becomes more and more uniform with increasing distance from the grating, since the contribution of the evanescence modes which have formed the scattered field decreases exponentially. At the distance $z \sim a$ (see curve 3 in Fig. 2) we deal with the case the far-field zone, where the main contribution to the transmitted wave originates from the mode with $m = 0$.

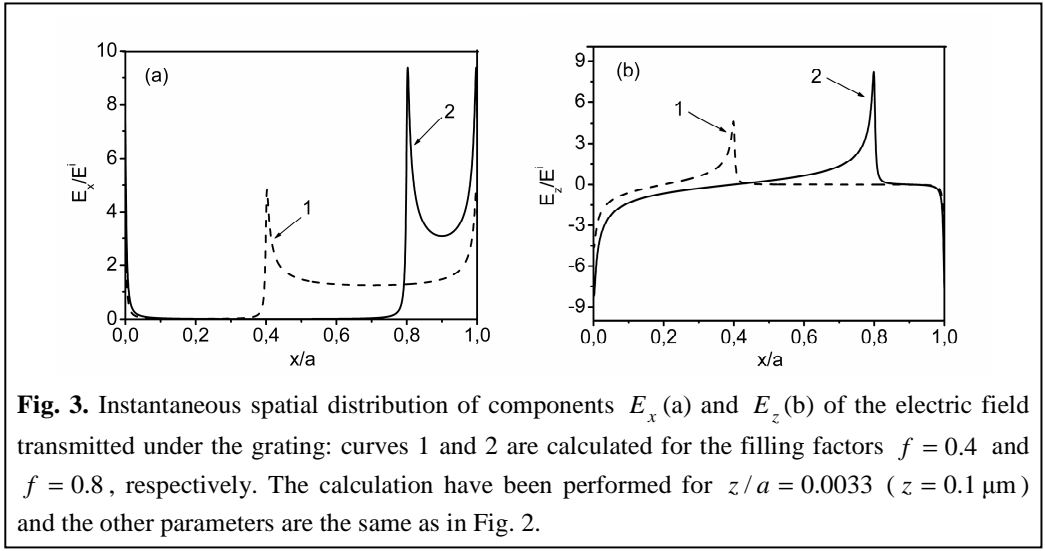

**Fig. 3.** Instantaneous spatial distribution of components $E_x$ (a) and $E_z$ (b) of the electric field transmitted under the grating: curves 1 and 2 are calculated for the filling factors $f = 0.4$ and $f = 0.8$, respectively. The calculation have been performed for $z/a = 0.0033$ ($z = 0.1\,\mu m$) and the other parameters are the same as in Fig. 2.

The distribution of the electric field in the near-field zone also depends on the grating parameters. As shown in Fig.3, the effect of enhancement of the electric field is more pronounced for the gratings with larger filling factors (i.e., narrower windows). For the case of $f = 0.8$ (see curve 2 in Fig. 3), the field distribution under the window is stipulated by the charges induced at the right and left edges of the adjacent stripes. Obviously, decreasing distance between these charges leads to increasing amplitudes of electric fields.

In order to clarify the major features of the near-field zone, Fig.4 depicts spatial mapping of the normalized time-averaged density of the electric energy, $W(x,z)$, of the transmitted wave. This quantity is calculated as: $W(x,z) = \left(|E_x(x,z)|^2 + |E_z(x,z)|^2\right)/E^{i2}$, where $E_x(x,z), E_z(x,z)$ are given by Eqs. (11).

The metallic grating produces a strong spatial redistribution of the energy of the (initially uniform) incident wave. Fig. 4 clearly demonstrates the existence of several zones with different energy concentration. For relatively small distance from the grating, the energy of transmitted wave is mainly concentrated near the edges of the strip. This corresponds to three *hot zones* (see white regions in Fig.4.). A *cold zone* is formed in the region under the metallic stripe (see black region in Fig.4). The lowest energy concentration correspond to the middle of the stripe where both x- and z-component of the field are close to zero. While moving away from the grating, the EM-wave penetrates under the metallic stripe and the energy is transferred from the hot zones into the cold one. An almost uniform distribution of the $W(x,z)$ parameter is established at the distance from the grating plane as large as $z \sim a/2$ (see panel Fig. 4a). When the grating has a structure with larger filling factor (see

Fig. 4b) we get stronger redistribution of the energy and therefore the near-field zone becomes larger.

Nowadays, there is number of experimental techniques that allow obtaining objects images beyond the diffraction limit, using special probes with sub-wavelength apertures [12]. Then a planar object scanned in front of this aperture could be imaged with a resolution determined by the aperture size rather than the radiation wavelength. A high-resolution technique for terahertz near-field imaging, which employs a planar structure with a single sub-wavelength aperture and 2D electron gas as detector, has recently been reported in Ref. [13]. In principle, the similar methods may be applied while visualizing the THz fields in the near-field zone of the periodic metallic structures.

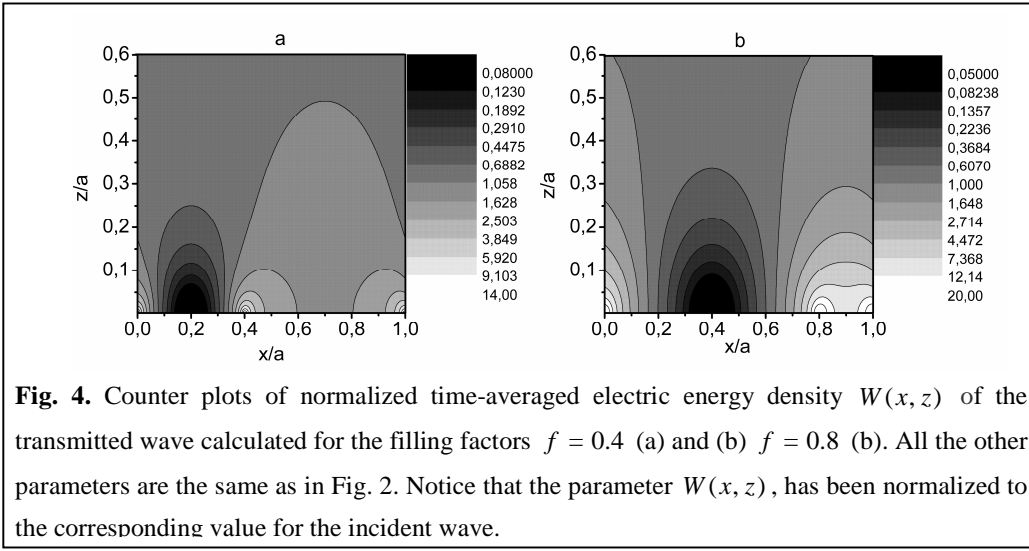

**Fig. 4.** Counter plots of normalized time-averaged electric energy density $W(x,z)$ of the transmitted wave calculated for the filling factors $f = 0.4$ (a) and (b) $f = 0.8$ (b). All the other parameters are the same as in Fig. 2. Notice that the parameter $W(x,z)$, has been normalized to the corresponding value for the incident wave.

### 3.1. Polarization of the electromagnetic wave in the near-field zone

Al already mentioned above, the real vector of the total electric field in the near-field zone has two components: $E_{x,z}(x,z,t) = \text{Re}[E_{x,z}(x,z)e^{-i\omega t}]$. During one oscillation period, a terminus of the local electric field vector will circumscribe some *polarization ellipse* in the $\{E_x, E_z\}$ plane. The ellipticity, $\delta$, and the azimuth angle $\alpha$ of the polarization ellipse may be expressed as follows:

$$\delta = \left[\left(1 - \sqrt{1-\beta^2}\right) \big/ \left(1 + \sqrt{1-\beta^2}\right)\right]^{1/2}, \tan\alpha = \left[2r^2 - (1+r^2)\left(1-\sqrt{1-\beta^2}\right)\right] \big/ 2r\cos(\Delta\phi),$$

where $\beta = 2r\sin(\Delta\varphi)/(1+r^2)$, $r = |E_z(x,z)|/|E_x(x,z)|$, $\Delta\phi$ denotes the phase shift between the components $E_x(x,z,t)$ and $E_z(x,z,t)$. It is evident, that these two polarization parameters obviously would depend upon the local point $(x,z)$.

The left panel of Fig. 5 illustrates the local polarization ellipses at fixed coordinate $z$ for different coordinates $x$. They correspond to the regions under the strip, near the edge of the strip and under the window. The right panel Fig. 5 depict evolution of the polarization ellipses with increasing of distance from the grating at fixed coordinate $x$. Both the $z$- and $x$- components tend to zero very closely to the middle of the stripe, and the polarization ellipse degenerates into a point. While moving towards the edge of the strip, the $z$-component is rapidly increases though the $x$-component remains small, ($r \gg 1$). Thus, we observe increase in both the length of the major axis of the ellipse and the azimuth angle $\alpha$ (see curves 1 and 2 in Fig. 5a). Large parameter $r$ also induce small ellipticities of the polarization ellipses. Near the edge of the stripe (see curve 3 in Fig. 5a), the $x$- and $z$-components oscillate practically in-phase ($\Delta\phi \sim 0$), with the close amplitudes ($r \sim 1$). As the consequence, the polarization ellipse degenerates into a straight-line, with $\alpha \sim 45°$. In the region under the window $r \ll 1$, we have the ellipse with a small azimuth angle and, again, a small ellipticity (see curve 4 in Fig. 5). The parameter $r$ decreases with increasing of distance from the grating and we observe a monotonic decrease in the $\alpha$ parameter (see Fig. 5b). The polarization ellipses degenerates into horizontal straight lines far from the grating, so that the transmitted EM-wave becomes linearly polarized.

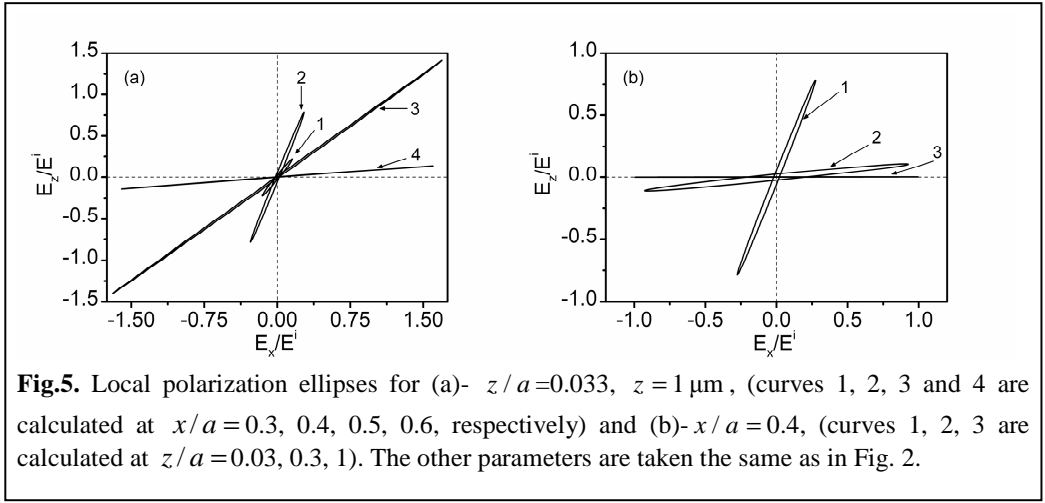

**Fig.5.** Local polarization ellipses for (a)- $z/a = 0.033$, $z = 1\,\mu m$, (curves 1, 2, 3 and 4 are calculated at $x/a = 0.3$, 0.4, 0.5, 0.6, respectively) and (b)- $x/a = 0.4$, (curves 1, 2, 3 are calculated at $z/a = 0.03$, 0.3, 1). The other parameters are taken the same as in Fig. 2.

Summarizing the results obtained in this section, we note that metallic grating produces a complicated wave field in near-field zone, which is characterized with strong spatial dependences of both the local energy and the polarization characteristics. Such a property in the near-field can be used for a selective THz photoexcitation of various types nanostructures, nanoobjects, as well as molecules, especially in case if excitation mechanisms are sensitive to both the polarization and the amplitudes of the EM-field.

## 4. Conclusion

We have presented the solution procedures of solving the problem of interaction of THz-range EM-wave and the metallic grating having sub-wavelength period. We have shown that EM-wave in the near-field zone under the metallic grating has a complicated vector structure, which is result of superposition of the two fields, one of them being a field of dipoles with the charges induced at the edges of the metallic strips, and the second representing a field of the incident plane wave. It has been demonstrated that the total electric field of the transmitted wave has the two components along the direction parallel to the axis of grating and perpendicular to the grating plane. In the near-field zone, the amplitude ratio and phase difference of these components vary depending on local point under the grating. We have derived time-averaged density of the electric energy for the spatial region under the grating. The effect of the strong concentration of electric energy of THz wave near the edges of the metallic strips has been found. Some application of the effects mentioned have been suggested for controlling THz-excitation of various nanostructures and nanodevices.

## 5. Acknowledgement